\documentclass[12pt]{iopart}
\usepackage[dvips]{graphicx}
\usepackage[dvips]{color}
\usepackage{iopams}
\usepackage{mhequ}

\definecolor{MyColor}{rgb}{0.0,0.4,1}

\def\L{{\mathbb{L}}}
\def\J{{\mathcal{J}}}

\def\e{\varepsilon}

\def\mX{\mathcal{X}}
\def\mV{\mathbf{v}}

\def\mY{\mathbf{y}}
\def\mZ{\mathbf{z}}
\def\mC{\mathbf{c}}

\def\mW{\mathbf{W}}

\let\rho=\varrho
\def\ddelta{\tilde{\delta}}

\newcommand{\cor}[1]{\langle #1\rangle}

\begin{document}

\title[Dynamics of anomalous heat transport: numerical analysis]
{Nonequilibrium dynamics of a stochastic model of anomalous heat
transport: numerical analysis}

\author{L Delfini $^{1}$\footnote{Present address:
Universit\'e de Toulouse, UPS, Laboratoire de Physique Th\'eorique 
(IRSAMC), F-31062 Toulouse, France}, S Lepri $^{1}$, R Livi $^{2,3}$, C
  Mej{\'{\i}}a-Monasterio $^{1}$\footnote{Present 
    address: University of Helsinki, Department of Mathematics and Statistics,
     P.O. Box 68 FIN-00014, Helsinki, Finland} and A Politi $^{1}$} 
\address{$^{1}$ Istituto dei Sistemi Complessi, Consiglio Nazionale
delle Ricerche, via Madonna del Piano 10, I-50019 Sesto Fiorentino, Italy
}
\address{$^{2}$ Dipartimento di Fisica, Universit\`a di Firenze, 
via G. Sansone 1 I-50019, Sesto Fiorentino, Italy 
}
\address{$^{3}$ Sezione INFN, and CSDC Firenze,
via G. Sansone 1 I-50019, Sesto Fiorentino, Italy } 

\begin{abstract}
  We  study heat  transport in  a chain  of harmonic  oscillators with
  random elastic collisions  between nearest-neighbours. The equations
  of motion of  the covariance matrix are numerically  solved for free
  and fixed boundary conditions.  In the thermodynamic limit, the shape
  of the temperature profile and the value of the
  stationary heat flux depend on the choice of boundary conditions.
  For free boundary conditions, they also depend on the coupling strength
   with the heat baths. Moreover, we find a strong violation of local
   equilibrium at the chain edges that determine two boundary layers of
   size $\sqrt{N}$  (where $N$ is the chain
  length),  that are  characterized by  a different scaling behaviour
  from the bulk. Finally, we investigate  the relaxation towards the
  stationary
  state, finding  two long time  scales: the first corresponds  to the
  relaxation of the hydrodynamic modes; the second is a
  manifestation of the finiteness of the system.
\end{abstract}

\section{Introduction}

The problem of  heat transport in chains of oscillators  is one of the
most  relevant   testing  grounds  to  understand   the  behaviour  of
statistical  systems steadily kept  out of  equilibrium.  In  the last
decade, numerical simulations  and analytic arguments have contributed
to clarify  the behaviour of  such systems in the  thermodynamic limit
(see review papers  \cite{BLR00,LLP03,DHAR09} and references therein).
However, there is still a number of open questions such as the role of
Boundary Conditions (BC in  the following) and the convergence towards
the stationary state.  In spite of the continuous increase of computer
performances, direct numerical simulations  are still not so effective
as to  provide reliable data  on sufficiently large systems.   In this
respect,  stochastic models  like the  one introduced  in \cite{BBO06}
prove  very helpful.   In this  paper we  consider a  version  of such
models already analyzed in \cite{DLLP08,LMMP09}. The model consists in
a chain  of $N$  coupled harmonic oscillators  in interaction  (at the
boundaries) with two stochastic  heat baths at different temperatures.
In addition, the oscillators are subject to stochastic collisions that
exchange  the  momenta  of   randomly  chosen  pairs  of  neighbouring
oscillators, so  that both  energy and momentum  are conserved.   In a
sense, collisions  simulate the presence  of nonlinear terms,  as they
contribute to ensuring ergodicity of an otherwise integrable model. In
fact,  it has  been observed  that this  model closely  reproduces the
behaviour of  standard nonlinear systems such as an FPU-$\beta$ chain,
starting   from    an   anomalous   (diverging)    heat   conductivity
\cite{DLLP08}.   On  the  other  hand,  being  the  collision  rule  a
perfectly  linear  process,   the  evolution  equations  for  ensemble
averages of the  relevant observables can be written  in an exact form
and  thereby  solved numerically,  without  having  to  deal with  the
statistical fluctuations that affect  finite samples.  As a result, we
have found that the  invariant measure can be effectively approximated
by   the  product   of  Gaussian   distributions  aligned   along  the
eigendirections of  the covariance matrix  \cite{DLLP08}. Moreover, in
\cite{LMMP09}, we investigated  the continuum limit (which corresponds
to the  large-$N$ limit) of  the covariance matrix,  deriving suitable
partial differential  equations for the stationary state,  in the case
of fixed BC.  As a result,  we have obtained explicit formulae for the
temperature profile  and the energy current.  Remarkably,  this is the
first example of an analytic expression for the temperature profile in
a system characterized by anomalous heat transport.

In \cite{partI} we go beyond, by  extending the continuum limit to include the
time  dependence of  the covariance  matrix. The  reader is  thus  referred to
\cite{partI}  for   a  more   detailed  introduction  and   the  corresponding
bibliography.  The aim  of this paper is to  complement the analysis contained
in \cite{partI}  with accurate  numerical studies of  finite samples  with the
goal  of clarifying  those issues  that  are too  difficult to  be worked  out
analytically. We start by  numerically computing the stationary covariance for
both free and fixed  BC. This helps to shed some light  on the nontrivial role
played by BC, whenever heat  transport exhibits an anomalous behaviour. In the
presence of normal transport, one expects that BC affect only a finite boundary
layer so that, in the thermodynamic limit the leading term of the heat flux is
independent of BC. On the other hand, it is known that in disordered chains of
linear  oscillators,   the  same   system  may  even   behave  as   a  thermal
superconductor or as  an insulator, by simply switching from  free to fixed BC
\cite{PRep}.  In generic nonlinear  chains, numerical simulations suggest that
the  heat flux  scales in  the  same way,  independently of  BC. However,  the
careful simulations performed in  \cite{Comm} revealed that in the FPU-$\beta$
model, the ratio  between the heat fluxes measured for free  and fixed BC does
not  converge to  1 for  $N\to\infty$. Here  we show  that the  same behaviour
occurs in  our stochastic model.  Actually the dependence  on BC is  even more
subtle than one could  have imagined: while in the case of  fixed BC, the heat
flux  and  the  temperature  profile  are asymptotically  independent  of  the
coupling strength with the thermal baths, the same is not true for free BC.

A second  objective of this paper  is the analysis of  the convergence towards
the  steady state.   This question,  which has  been hardly  discussed  in the
literature, can be straightforwardly addressed for our model, as it amounts to
computing  the eigenvalues  of  the evolution  operator  for the  covariances.
Moreover,  we  also compare  the  convergence of  the  average  heat flux  for
different system sizes to show how careful direct simulations must be, if they
have  to be  trusted. We  find that finite--size  effects
associated  with the relaxation  rates of  slow, i.e. long-wavelength, modes
significantly  modify the  asymptotic scaling  of the  relaxation  process. In
practice, we find numerical evidence that the theoretical hydrodynamic
scaling holds only over a finite range of time scales, although its duration
diverges with $N$. 

The paper is organized as follows. In Section \ref{sec:cov}, we briefly recall
the definition of the covariance matrix, and the coupled equations governing
its evolution towards the stationary value. Some  properties of  the steady
state are discussed in Section \ref{sec:stat}. The problem of the approach to
the  steady  state  is  addressed  in  Section  \ref{sec:relax}.  Finally,  in
Section~\ref{sec:concl} we summarize our main results.

\section{Equations for the covariance matrix}
\label{sec:cov}

In this section, we introduce  the minimal notations and definitions needed to
follow the  main discussion presented  in the following sections.   The reader
interested in  a more  detailed presentation is  referred to  \cite{partI}. We
consider a  chain of $N$ unit-mass particles  interacting via nearest-neighbour
harmonic coupling of frequency $\omega$. The equations of motion are given by
\begin{equa}[2]
\label{eq:eqs-motion}
& \dot q_n &  =  & p_n \\
& \dot p_n &  =  & \omega^2 (\ddelta_{n,N}q_{n+1} - 2q_n +
\ddelta_{n,1}q_{n-1}) + \delta_{n,1}(\xi^+ - \lambda \dot q_1) +
\delta_{n,N}(\xi^- -\lambda \dot q_N)  \ ,
\end{equa}
where $p_n$, $q_n$ are the momentum and displacement from equilibrium position
of   the  $n$-th  particle,   $\delta_{i,j}$  is   the  Kronecker   delta  and
$\ddelta_{i,j}  \equiv 1-\delta_{i,j}$, and  $\xi^\pm$ are  independent Wiener
processes with zero mean and variance $2\lambda k_B T_\pm$, where $k_B$ is the
Boltzmann constant and  $\lambda$ is the coupling constant.   In the following
free and fixed boundary conditions  will be considered. These can be expressed
in terms  of the position variable  $q$ as: $q_0=q_1$,  $q_N=q_{N+1}$ for {\it
  free} BC and $q_0=q_{N+1}=0$ for {\it fixed} BC.

We consider the covariance matrix written as
\begin{equation} 
\label{eq:covmat}
\mC = \left(
\begin{array}{cc}
  \mY       &   \mZ\\
  \mZ^\dag  &   \mV
\end{array}
\right) \ .
\end{equation}
where  the   matrices  $\mY$,  $\mZ$   and  $\mV$,  of   respective  dimension
$(N-1)\!\times\!(N-1)$, $(N-1)\!\times\!N$ and $N\!\times\!N$ are defined as
\begin{equation} \label{eq:defcov}
\mY_{i,j} = \cor{\Delta q_i \Delta q_j} \ , \\
\mZ_{i,j}  =  \cor{\Delta q_i p_j} \ , \\
\mV_{i,j}  =  \cor{p_i p_j} \ , \\
\end{equation}
where $\langle\cdot\rangle$  denotes the average over  phase space probability
distribution  function $P$  and $\Delta  q_i =  q_i -  q_{i-1}$ stand  for the
particle  relative displacements.   The variables  $(\Delta q_i,p_i)$  are the
convenient choice to deal with: one the one hand, absolute positions are not
well
defined for free  BC and on the other hand, the  potential energy is expressed
in terms  of relative  differences. The only subtlety is  that the  domain of
definition of  $\Delta q_i$ differs from  that of $p_i$. Thereby,  the bulk of
the system is defined as $\{\Delta q_i \ | \ i \in [2,N]\}$ and $\{p_i \ | \ i
\in [2,N-1]\}$.  The evolution equations for $\mC$ in the bulk are
\begin{equa}[2] \label{eq:Cdot}
& \dot \mY_{i,j} & \ = \ & \mZ_{j,i}-\mZ_{j,i-1}+\mZ_{i,j}-\mZ_{i,j-1} \ , \\ 
& \dot \mZ_{i,j} & \ = \ & \mV_{i,j}-\mV_{i-1,j}+
\omega^2\left(\mY_{i,j+1}-\mY_{i,j}\right)+
\gamma\left(\mZ_{i,j+1}+\mZ_{i,j-1}-2\mZ_{i,j}\right)\ ,\\
& \dot \mV_{i,j} & \ = \ & \omega^2\left(\mZ_{j+1,i}-\mZ_{j,i}+
\mZ_{i+1,j}-\mZ_{i,j}\right) + \gamma\mW_{i,j}  \ .
\end{equa}
These equations follow from the deterministic equations of motion
\eref{eq:eqs-motion} plus the contribution of the stochastic noise
($\gamma$ denotes the collision rate), that is described
by the collision matrix $\mW$,
\begin{equation} \label{wmat}
\mW_{ij} \equiv \left\{
\begin{array}{ll}
\mV_{i-1,j-1} + \mV_{i+1,j+1} - 2\mV_{i,j} & i = j  \\ 
\mV_{i-1,j} + \mV_{i,j+1} - 2\mV_{i,j} & i - j = -1 \\
\mV_{i+1,j} + \mV_{i,j-1} - 2\mV_{i,j} & i - j = 1  \\
\mV_{i+1,j} + \mV_{i-1,j} + \mV_{i,j-1} + \mV_{i,j+1}
- 4\mV_{i,j} & |i - j| > 1
\end{array}
\right. \ .
\end{equation}
On the boundaries, several changes appear in the velocity fields. The
interested reader can find a full description in Section 2.2 of
\cite{partI}.  Here we limit ourselves to show the contribution
arising from the coupling with the heat bath, namely
\begin{eqnarray}
 \dot \mZ^b_{i,j} &=& \delta_{j,1}\mZ_{i,1} + \delta_{j,N}\mZ_{i,N} \ ,
 \nonumber \\
 \dot \mV^b_{i,j} &=& \delta_{j,1}\mV_{i,1} + \delta_{j,N}\mV_{i,N} + \delta_{i,1}\mV_{1,j} +
\delta_{i,N}\mV_{N,j} - 2\left(T^+\delta_{i,1}\delta_{j,1} +
T^-\delta_{i,N}\delta_{j,N}\right) \ . \nonumber
\end{eqnarray}

\section{Stationary covariance}
\label{sec:stat}

In this  section we investigate  some properties of the  nonequilibrium steady
state,  for both  fixed and  free  BC.  The  stationary state  is obtained  by
considering the time-independent solution  of equations \eref{eq:Cdot}. It can
be  efficiently determined  by exploiting  the sparsity  of  the corresponding
linear problem, as  well as the symmetries of the  unknowns (this approach has
been followed in \cite{DLLP08} for  fixed BC). Alternatively, one can just let
evolve   equations  \eref{eq:Cdot}  starting   from  any   meaningful  initial
conditions, as the dynamics will  necessarily converge towards the only stable
stationary state  (here we have adopted  this latter approach  also because we
wish to study the convergence -- see in the following).  All numerical results
presented  in  this  paper  have  been  obtained  for  $\omega=1$,  $T_+=1.5$,
$T_-=0.5$.   This  is by no means  a  limitation, as  all  these
parameters  can be  easily  scaled out  due  to the  linear  structure of  the
model. Accordingly, they will not be mentioned again, unless specifically needed
for a comparison with theoretical predictions.

\subsection{The heat flux}

The first  observable we  have looked at  is the  energy flux at  position $i$
which, in terms of the matrices $\mV$, $\mZ$ is written as \cite{PRep}
\begin{equation}
J_i = -\omega^2\mZ_{i+1,i+1} + \frac{\gamma}{2}(\mV_{i,i}-\mV_{i+1,i+1}) . 
\end{equation}
We  have  adopted  the  convention  that  a positive  flux  corresponds  to  a
propagation towards increasing values of the spatial index $i$. The first term
stems from the  deterministic forces and provides for  the leading (anomalous)
contribution,  while the  second  one  accounts for  energy  exchanges due  to
collisions of nearby particles. In  the stationary state, $J_i$ is independent
of $i$, i.e. $J_i \equiv J$ .

\begin{figure}[!t]
\begin{center}
\includegraphics[scale=0.45,clip]{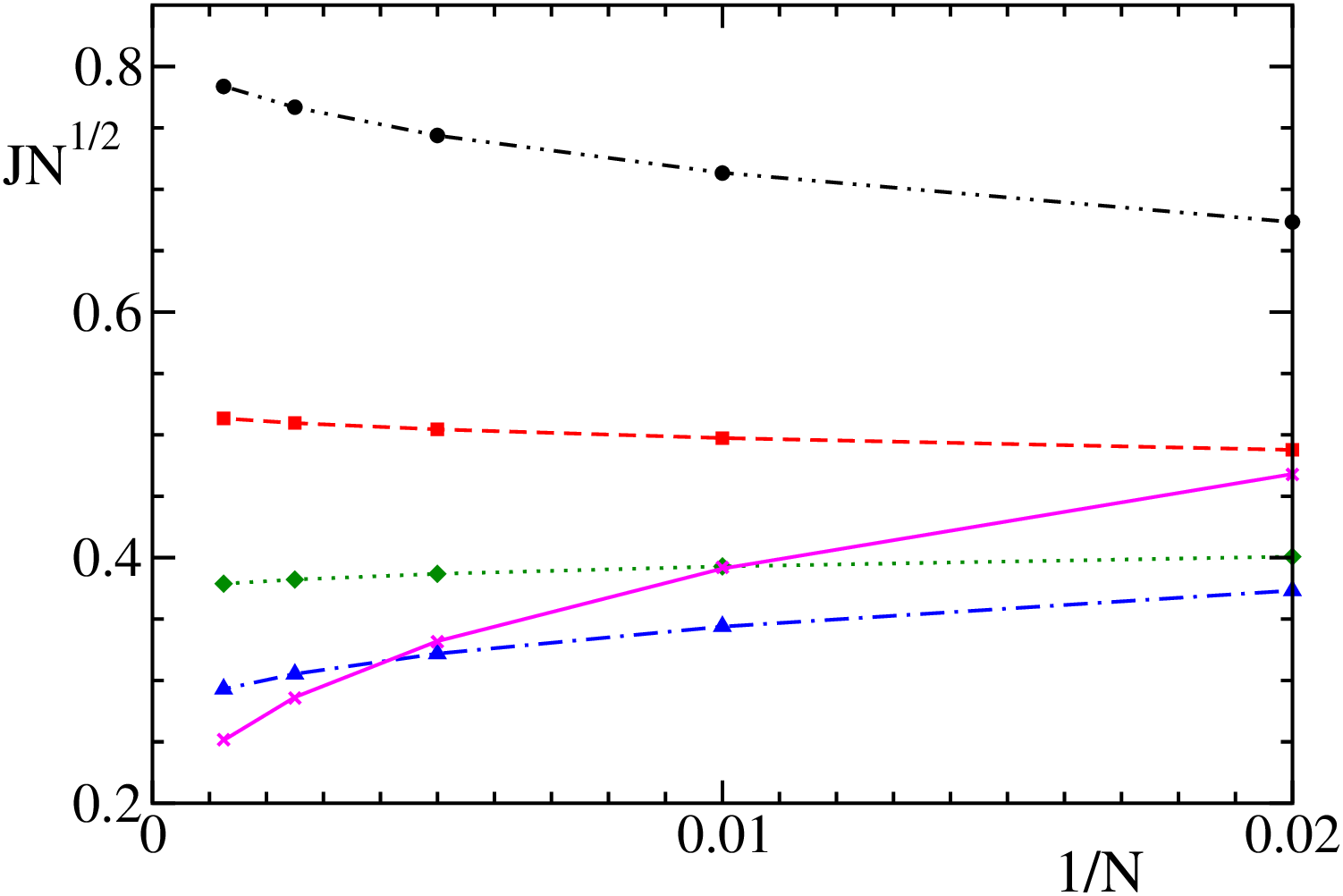}
\caption{Free BC ($\lambda=1$): the  scaled stationary flux $J\sqrt{N}$ versus
  $1/N$   for  $\gamma$=0.2   (circles),  0.5   (squares),  1   (diamonds),  2
  (triangles), 5 (crosses).   The lines are obtained by  fitting the data with
  equation~\eref{eq:fit}.}
\label{fig1}
\end{center}
\end{figure}

In figure \ref{fig1}  we show $J\sqrt{N}$ as a function of  the inverse of the
system size  $N$. The results  refer to  free BC (as  we do not  have analytic
estimates to compare with), $\lambda=1$  and different values of the collision
rate $\gamma$  (see the various symbols  as described in  the figure caption).
In all cases there is a convincing evidence that $J \sim N^{-1/2}$, similar to
what predicted analytically  in \cite{LMMP09} for the case of  fixed BC.  As a
consequence, the effective conductivity, $\kappa \equiv JN/(T_+-T_-)$ diverges
as  $\sqrt{N}$.   However, from  figure  \ref{fig1}  it  is also  evident  the
presence of sub-leading singular corrections which hinder the extrapolation of
the asymptotic value. In analogy to \cite{LMMP09}, we introduce the Ansatz,
\begin{equation}
\label{eq:fit}
J\;=\;\frac{\J}{\sqrt{N}}+\frac{B}{N^{\beta}}
\end{equation}
By using this formula to fit the data, we obtain the curves reported in figure
\ref{fig1}  which  reproduce  quite  well  the  raw  data.   Notice  that  the
convergence is  from below for smaller  $\gamma$ values, while  from above for
larger collision rates. All the estimated $\beta$ values range in the interval
$[0.88,0.95]$, suggesting that this parameter may be ``universal''.

\begin{figure}[!t]
\begin{center}
\includegraphics[scale=0.45,clip]{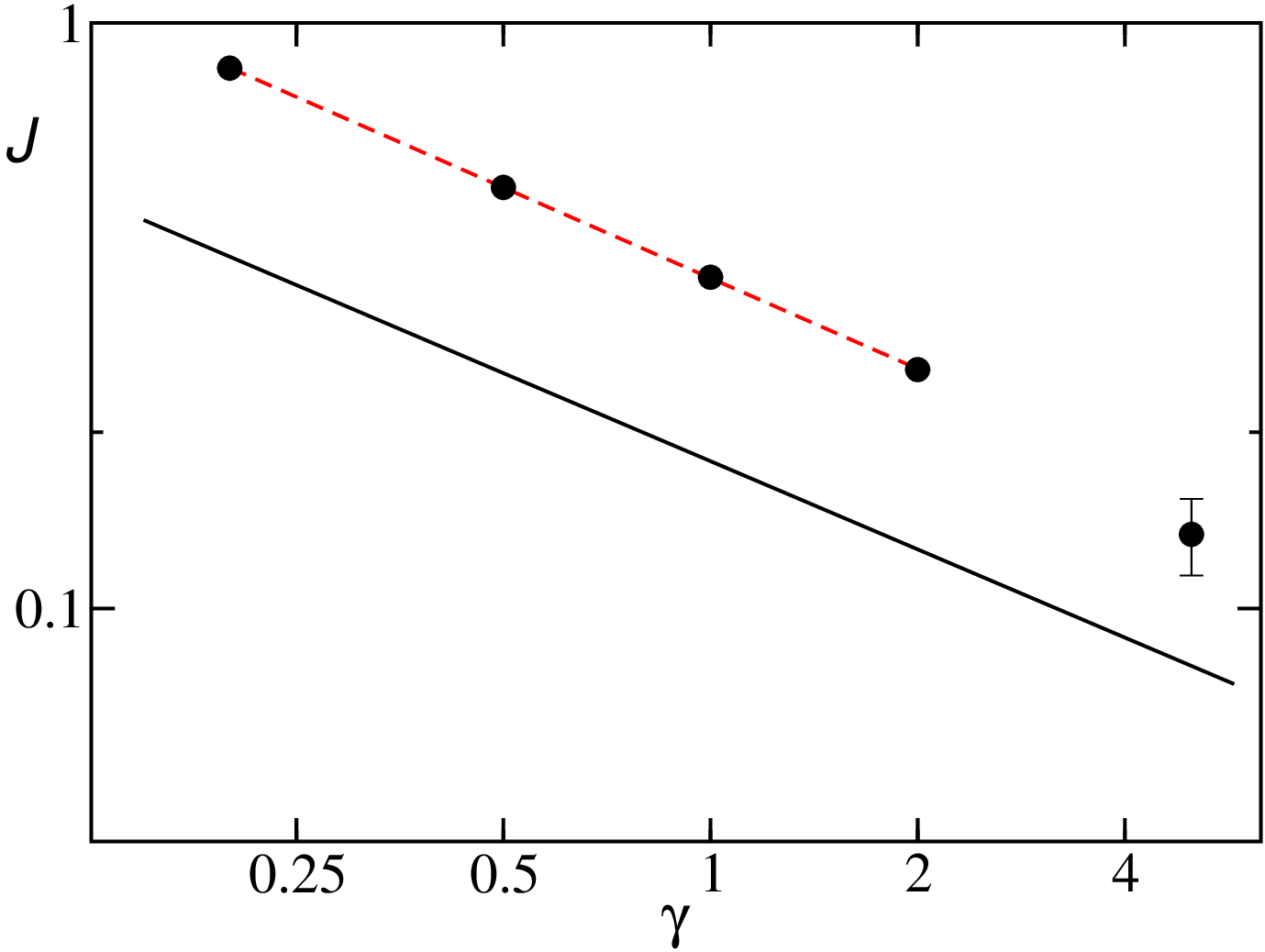}
\caption{Asymptotic value of $\J$ as a function of the collision rate $\gamma$
  for free BC and the same  parameter values as in figure \ref{fig1}. Data are
  plotted in log-log scales.  The  error bar for $\gamma=5$ has been estimated
  from a  rough comparison among different extrapolation  schemes.  The dashed
  line is  a power law fit  of the first four  data: its slope  is -0.51.  The
  solid line corresponds to the  analytic solution for fixed BC (equation (20)
  of \cite{LMMP09}). }
\label{fig2}
\end{center}
\end{figure}

The extrapolated $\J$  values are plotted in figure  ~\ref{fig2}, where we can
see that $\J  \sim \gamma^{-1/2}$, as found for fixed  BC \cite{LMMP09} .  For
$\gamma=5$, the  extrapolated value of $\J$ suffers  a substantial uncertainty
due to large  finite-size corrections (that become even  more sizeable for yet
larger $\gamma$ values).

So far, we  have not found any relevant difference between  fixed and free BC.
The heat  flux scales  in the  same way in  both cases  and exhibits  the same
dependence on the collision rate. If the effect of the BC were restricted to a
layer of  finite width  around the boundary,  in the thermodynamic  limit, the
thermal  resistance of  a given  chain  would be  independent of  the type  of
thermal contact.  In  other words, we should expect $\J$  to be independent of
the BC.   However, this is  not the  case, as it  can be inferred  from figure
~\ref{fig2}, where  we have also plotted  the analytic curve for  the fixed BC
case  (equation  (20)  in \cite{LMMP09}).   For  free  BC,  the heat  flux  is
approximately twice  as that  obtained for fixed  BC. It is  worth mentioning
that  the same  effect  was found  in  the simulations  of FPU-$\beta$  chains
\cite{Comm}, although with a slightly different value of the ratio (around 1.7
in that  case).  Since the flux is  constant along the chain,  this means that
even deeply in the bulk, the system perceives the effect of the boundaries. In
particular, from the  knowledge of the local temperature  profile and from the
heat flux, one  can in principle infer the type of  BC.  These results suggest
that this is another way anomalous conduction manifests itself.
 
\begin{figure}[!t]
\begin{center}
\includegraphics[scale=0.45,clip]{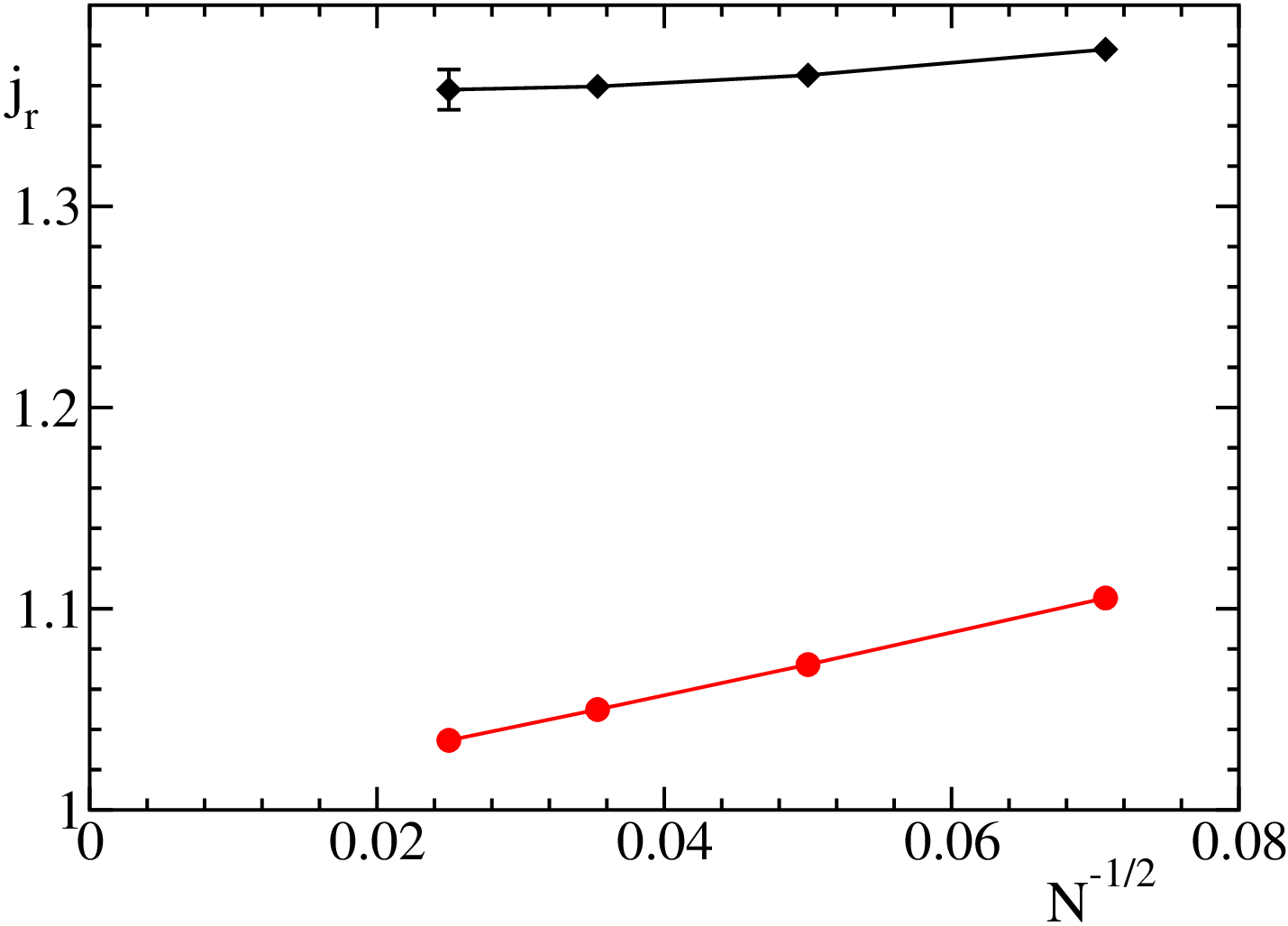}
\caption{Ratio of  the energy fluxes as  defined in equation~(\ref{eq:ratio}),
  for free  (diamonds) and fixed (circles)  BC as a  function of $1/\sqrt{N}$,
  and for $\gamma=1$. The error on $j_r$  for free BC and $N=1600$ is due to a
  not yet relaxed dynamics.}
\label{figrat}
\end{center}
\end{figure}

The whole scenario is even more subtle than suggested by figure \ref{fig2}. In
fact, for  free BC,  the leading  term of the  heat flux  depends not  only on
$\gamma$ but also on the coupling strength $\lambda$ with the heat bath, while
this is not so for fixed BC.  We illustrate this in figure \ref{figrat}, where
we plot the ratio
\begin{equation}
\label{eq:ratio}
j_r = j(\lambda=1,N)/j(\lambda=1/4,N)
\end{equation}
where $j(\lambda,N)$ is the heat flux in a chain of length $N$ and for a given
value of  $\lambda$. It is  not surprising to  see that the coupling  with the
heat baths modifies the flux in  chains of finite length. However, we see that
for fixed BC, the effect of the coupling vanishes as $j_r$ converges to 1 (see
the lower curve in figure ~\ref{figrat}).  On the contrary, for free BC, $j_r$
remains significantly different  from 1. This suggests that  fixed BC may lead
to  a  kind  of  universal  behaviour,  namely the  heat  flux  and  also  the
temperature  profile   are  independent  of  the  details   of  the coupling
with the heat baths. This is not the case for free BC. It would be
interesting to check whether the same holds true in generic nonlinear chains.

\subsection{The temperature profile}

\begin{figure}[!t]
\begin{center}
\includegraphics[scale=0.45,clip]{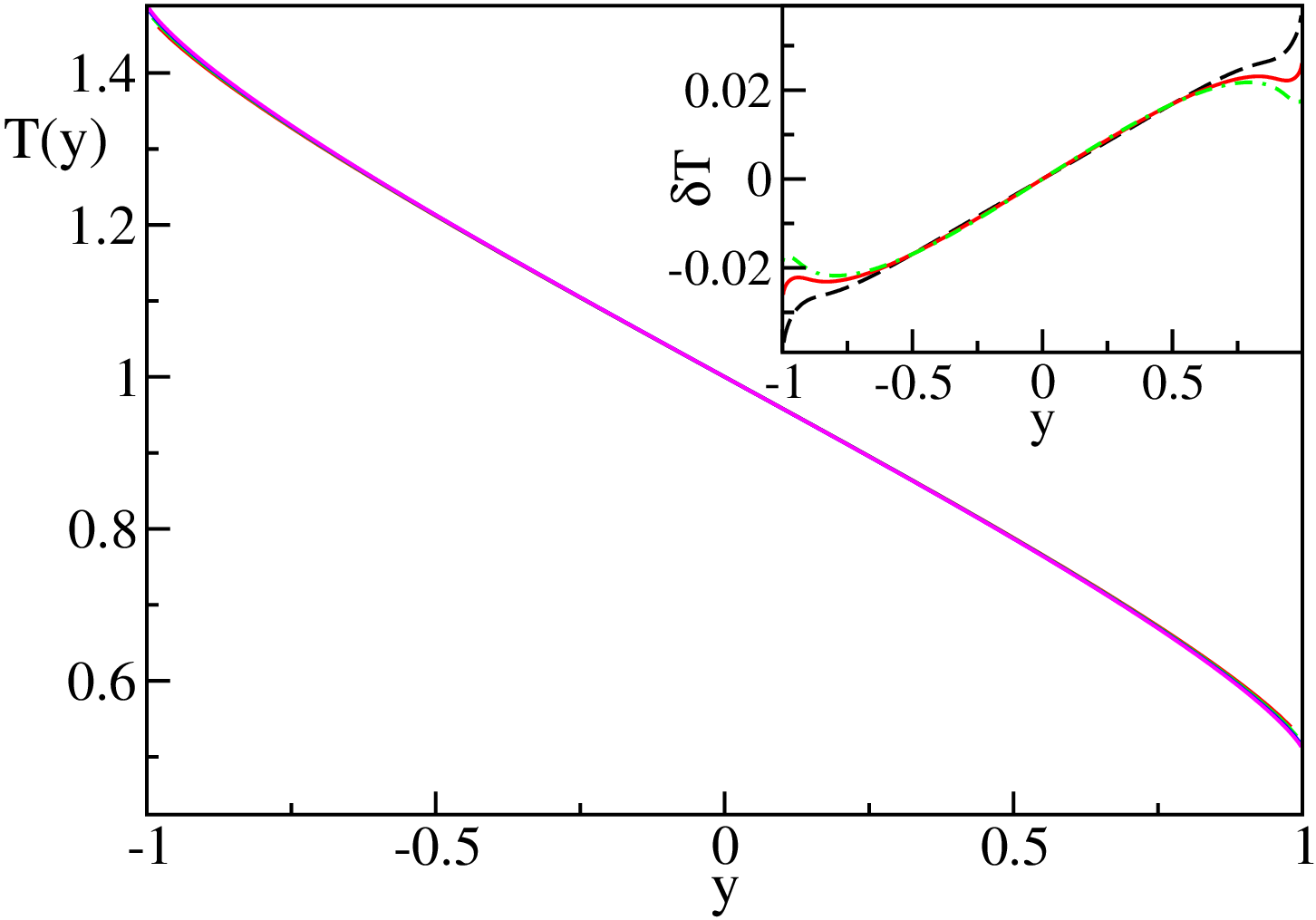}
\caption{Temperature profile  $T(y)$ for  $\gamma=1$ and $\lambda=1$  and free
  BC, for  $N$ = $100$,  $200$, $400$ and  $800$.  In the inset:  the relative
  difference between the temperature profiles corresponding to $\lambda=1$ and
  $\lambda=1/4$, with  $\gamma=1$ and for $N=400$, $800$,  and $1600$ (dashed,
  solid, and dotted-dashed lines, respectively).}
\label{profili}
\end{center}
\end{figure}

Another observable of interest is the temperature profile $T_i=\langle
p^2_i\rangle$.  In figure \ref{profili} we show the temperature
profile for free BC and three different sizes, as a function of the
``normalized'' position along the chain $y \equiv 2i/N -1$, that
varies in the interval $[-1,1]$. The data collapse is coherent with
the scaling assumed in the continuum approach \cite{LMMP09}.  The
shape of the profile is qualitatively similar to that obtained for
fixed BC but, although here there are no (square-root) singularities
at the boundaries of the chain (see equation (19) and (79) in
\cite{LMMP09}).  Furthermore, we notice that the profile itself
depends on both $\gamma$ and $\lambda$.  Evidence of such a dependence
can be appreciated in the the inset of figure \ref{profili}, where the
difference $\delta T(y)$ between the profiles corresponding to
$\lambda=1$ and $1/4$ is plotted in for three different system sizes.
In fact, we see that $\delta T(y)$ does not vanish in the
thermodynamic limit.  Moreover, the regions around the boundaries are
affected by strong finite-size effects. In fact, one expects that
$\delta T(-1)=\delta T(1)= 0$, as the temperature necessarily
converges, as $N\rightarrow\infty$, to that of the attached heat bath.

\subsection{Other correlators}

In  this  section  we  analyse  the behaviour  of  the  different  correlators
\eref{eq:defcov}, along  the diagonal ($i=j$) and  for generic values  of $x =
(i-j)/\sqrt{N}$.\footnote{The continuous coordinate  $x$ measures the distance
  of a given  correlator from the diagonal. We have used  the same notation in
  \cite{LMMP09}.} We first analyse the case of fixed BC.

\begin{figure}[!t]
\begin{center}
\includegraphics[scale=0.5]{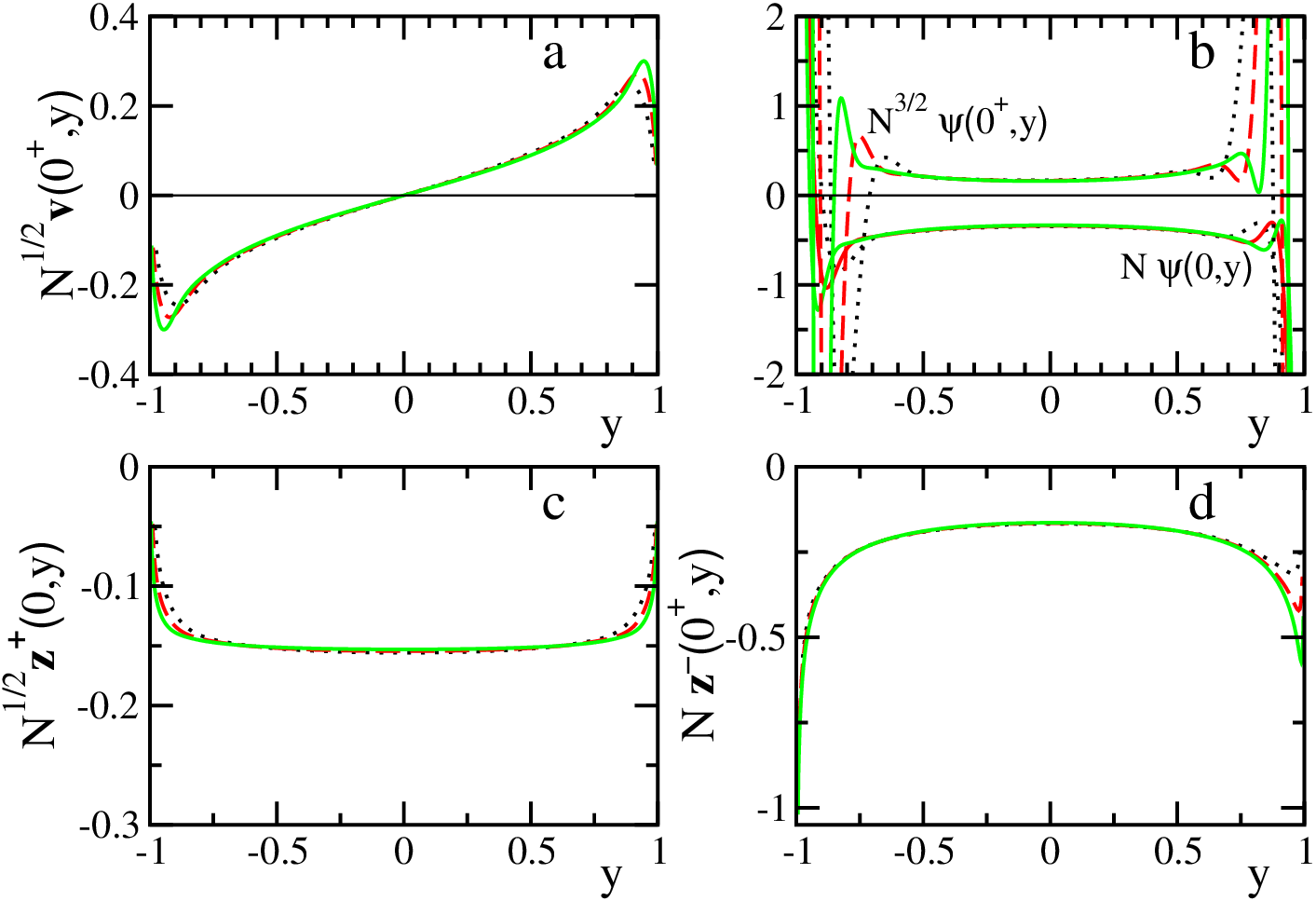}
\caption{Behaviour  of  the  stationary  covariances  as  a  function  of  the
  coordinate $y$ for fixed BC and different sizes: (a) First lower diagonal of
  $\mV$ (b)  diagonal (lower curves)  and first subdiagonal (upper  curves) of
  $\psi$  (c)  diagonal of  the  symmetric  component  $\mZ^+$ and  (d)  First
  subdiagonal of the antisymmetric  component $\mZ^-$.  In all panels, dotted,
  dashed and solid lines refer to $N=200$, $400$, and $800$ respectively.  The
  physical parameters are $\gamma=1$, $\lambda=1$.}
\label{covfx}
\end{center}
\end{figure}

At equilibrium, off-diagonal elements  of the correlators \eref{eq:defcov} are
zero.   In the nonequilibrium  steady state  we have  recently shown  that the
off-diagonal correlators $\mV$ are of $\Or(1/\sqrt{N})$ \cite{LMMP09}. This is
confirmed  in  figure  \ref{covfx}a,  where  we plot  the  lower  diagonal  of
$\mV_{i,i+1}$, corresponding  to $\mV$  measured at a  distance $x=1/\sqrt{N}$
from the diagonal, that is denoted by $x=0^+$.

The  potential energy  profile  $\mY$ closely  reproduces  the kinetic  energy
profile (see also \cite{DLLP08}). In  order to appreciate its contribution, it
is  necessary  to  look at  higher-order  corrections.  This  can be  done  by
introducing
\begin{equation}
\label{eq:psi}
\psi_{i,j} = \mV_{i,j} - \omega^2 \mY_{i,j}
\end{equation}
which measures the mismatch between kinetic and potential energy.
Figure \ref{covfx}b shows that along the diagonal (see the lower set
of curves), $\psi$ scales as $1/N$ everywhere except perhaps at the
boundaries.  From a physical point of view, this implies that
everywhere in the bulk, the system is locally at equilibrium (with
$1/N$ finite-size deviations from the virial equality).  The wild
behaviour observed near the boundaries suggests the existence of
nontrivial boundary layers. We will discuss this in detail in the next
subsection.  Analogously to $\mV$, $\psi$ exhibits a ``discontinuity''
when moving away from the diagonal.  Indeed, in the upper set of
curves of figure \ref{covfx}b we show that $\psi(x=0^+,y)$ is of order
$1/N^{3/2}$.  It is worthwhile remarking that this scaling holds only
for $\lambda=1$.  For$\lambda\ne1$ we have found that the off-diagonal
terms of $\psi$ are of order $1/N$ too. By recalling that we have here
selected $\omega=1$ , it is reasonable to conjecture that the faster
convergence of $\psi$ observed for $\lambda=1$ is a manifestation of
the thermal impedance matching on the boundary, theoretically
predicted for $\lambda=\omega$ (see \cite{partI}).

\begin{figure}[!ht]
\begin{center}
\includegraphics[scale=0.5,clip]{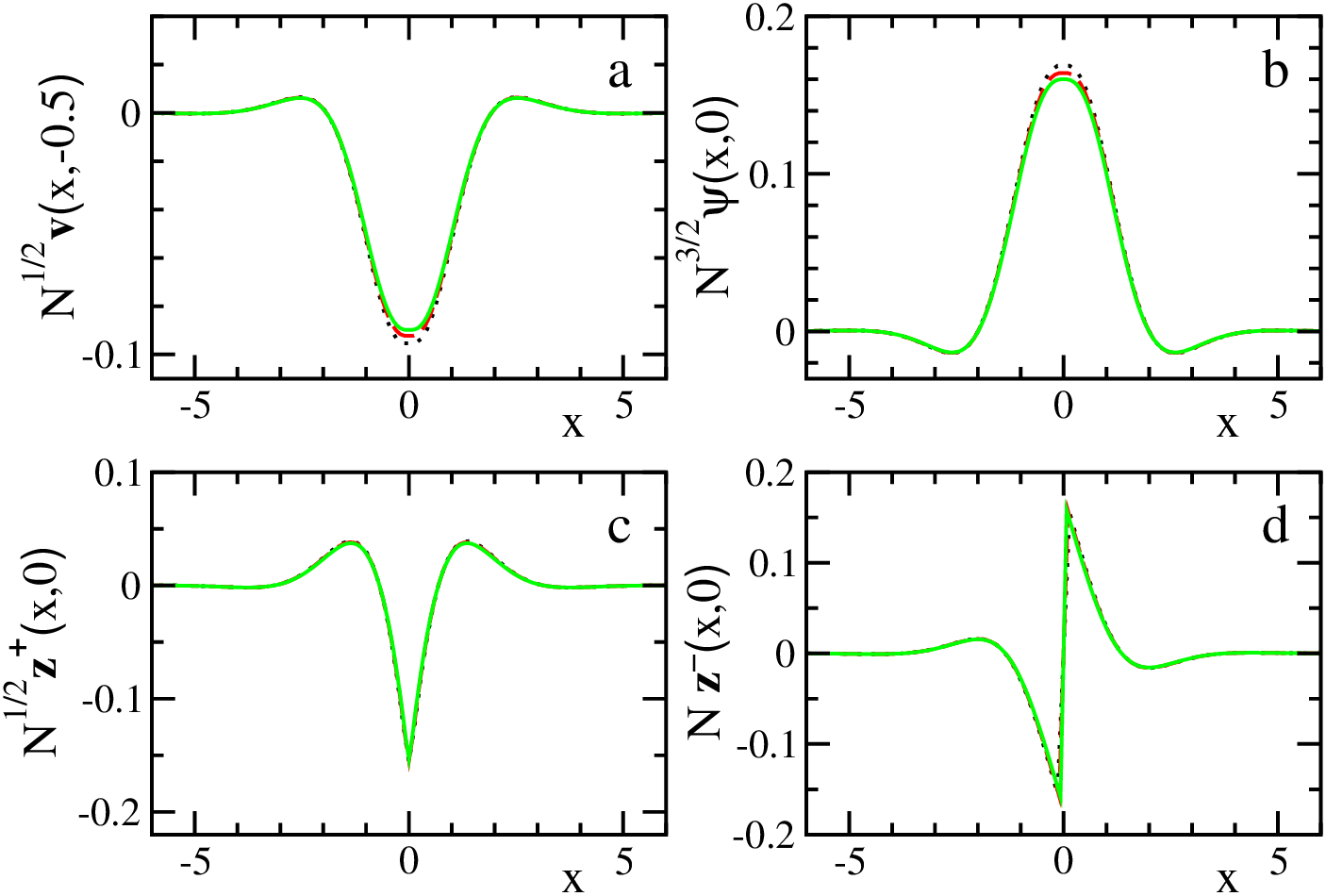}
\caption{Off-diagonal behaviour of the stationary covariances for fixed BC and
  the same parameter values and same  notations as in the previous figure: (a)
  $\mV$ for $y=-0.5$; (b) $\psi$ for  $y=0$; (c) $\mZ^+$ for $y=0$ (d) $\mZ^-$
  for $y=0$. The  physical parameters are $\gamma=1$ and  $\lambda=1$. For the
  sake of clarity,  in panels a and b the diagonal  values ($x=0$) are omitted
  as they are of lower order.}
\label{covfx2}
\end{center}
\end{figure}

Moreover, as  shown in \cite{partI},  it is convenient to  distinguish between
symmetric and anti-symmetric components of the correlators $\mZ$ with respect
to $x$,
\begin{equation}
\label{eq:zpm}
\mZ^{\pm}_{i,j} = \frac{\mZ_{i,j} \pm \mZ_{j,i}}{2} \ .
\end{equation}
In figure \ref{covfx}c,  we plot the symmetric component  which corresponds to
the leading term  of the heat flux.  In fact, it  scales as $1/\sqrt{N}$.  The
deviations from a perfectly flat shape reveal again the presence of nontrivial
boundary  layers. Along the  diagonal the antisymmetric component $\mZ^-$ is
zero by construction, while along the first subdiagonal, $\mZ^-$ scales as $1/N$ 
(see figure \ref{covfx}d).

In figure \ref{covfx2} we show the behaviour of the correlators as a function
of their distance $x$ from the diagonal. We have found that $\mZ^-$  and the
derivative of $\mZ^+$ along $x$ are both discontinuous across the diagonal, in
agreement with the  theoretical analysis in \cite{partI}. All  the results are
independent of  $\lambda$ except for  the variable $\psi$ which,  for $\lambda
\ne 1$, is constant away from the diagonal and of order $1/N$.  We would like
to remark that this anomaly does not affect the theoretical analysis carried in
\cite{LMMP09} and \cite{partI}, as (for fixed BC) $\psi$ does not contribute
to the leading behaviour of the temperature profile and of the heat flux.

The very  good overlap  among the curves  obtained for different  system sizes
confirms the scaling behaviour  of the off-diagonal  already  seen in
figure \ref{covfx}.  Most important, the observed scaling  corroborates the
validity  of the {\it   ansatz}  used in  \cite{LMMP09}  and  \cite{partI}.
Summarizing, for  fixed BC and  far from the boundary we find that:
\begin{itemize}
\item[--] Along the diagonal, $\psi$ is $\Or(1/N)$. Off-diagonal, $\psi$ is
  $\Or(1/N^{3/2})$ for $\lambda=\omega$ and $\Or(1/N)$ otherwise.
\item[--] The correlator $\mV$ is $\Or(1)$ along the diagonal and
  $\Or(1/\sqrt{N})$ off-diagonal.
\item[--] The symmetric correlator $\mZ^+$ is $\Or(1/\sqrt{N})$ everywhere.
\item[--] The antisymmetric correlator $\mZ^-$ is $\Or(1/N)$.
\end{itemize}

As a final  remark, note that $\mZ^+$ is the only  variable that is continuous
in $x$. This implies that  the difference $\mZ^+(0^+,y) - \mZ^+(0,y)$, that we
have denoted by $\delta\mZ^+$ in \cite{partI}, must necessary be an order $\e$
higher than its addenda, since its leading contribution  is a derivative with
respect to $x$.

\begin{figure}[!t]
\begin{center}
\includegraphics[scale=0.5,clip]{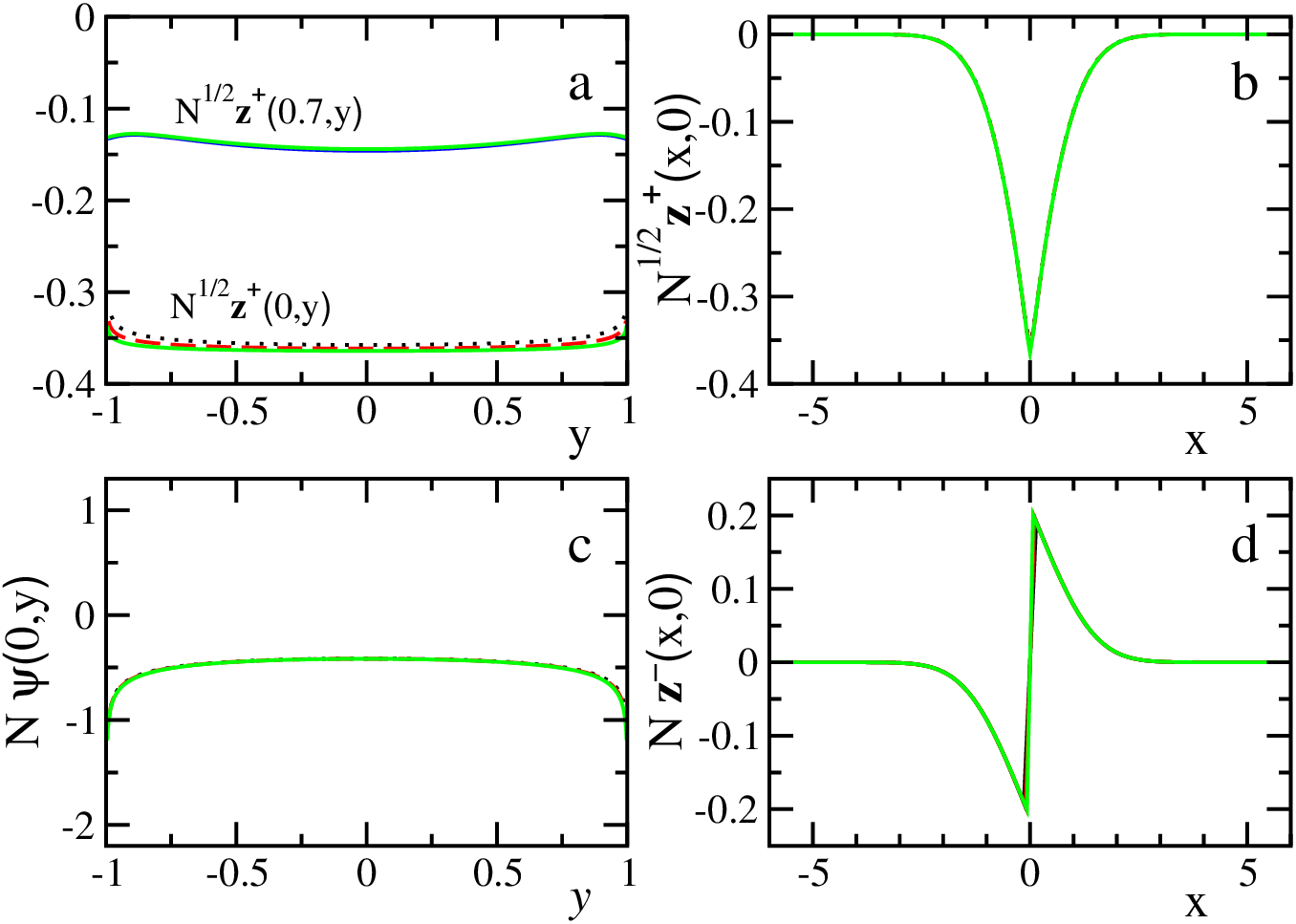}
\caption{Elements  of the  stationary covariances  for free  BC  and different
  sizes:  (a) $\mZ^+(x=0,y)$  (lower  curves) and  at $\mZ^+(x=0.7,y)$  (upper
  curves); (b)  $\mZ^+(x,y=0)$; (c) $\psi(x=0,y)$ (d)  $\mZ^-(x,y=0)$.  In all
  panels, dotted,  dashed and solid lines  refer to $N=200$,  $400$, and $800$
  respectively.}
\label{covfr}
\end{center}
\end{figure}

We  now  turn our attention  to  the  free  BC.  As it can be seen  in  figure
\ref{covfr}, the correlators scale with $N$ in the same manner, irrespectively
of  the  boundary  conditions.   We  only  notice  the  following  qualitative
differences: First,  with free  BC, the convergence at the boundaries  is more
effective  than  for  fixed   BC  (compare  figure  \ref{covfx}b  with  figure
\ref{covfr}c). Second, as  a function of $x$, some  additional oscillations of
$\mZ$  can  be  seen  only  for  fixed  BC  (figure  \ref{covfx}c  and  figure
\ref{covfr}b). Third,  $\mZ^+$ is  larger for free  BC, in agreement  with the
fact that in this case, the heat flux is about two times larger than for
fixed BC.

\subsection{Behaviour at the chain edges}

The numerical discussion carried out in the previous subsection has
revealed the existence of ``boundary layers'' in the vicinity of the
contact points with the heat baths ($y\approx\pm1$), where strong
deviations from the expected scaling behaviour are clearly visible.
Since in \cite{partI}, we have not attempted a theoretical analysis of
the boundary layers, it is at least necessary to clarify their
relevance, with reference to the numerical but otherwise exact
solutions for the correlators.

\begin{figure}[!t]
\begin{center}
\includegraphics[scale=0.45,clip]{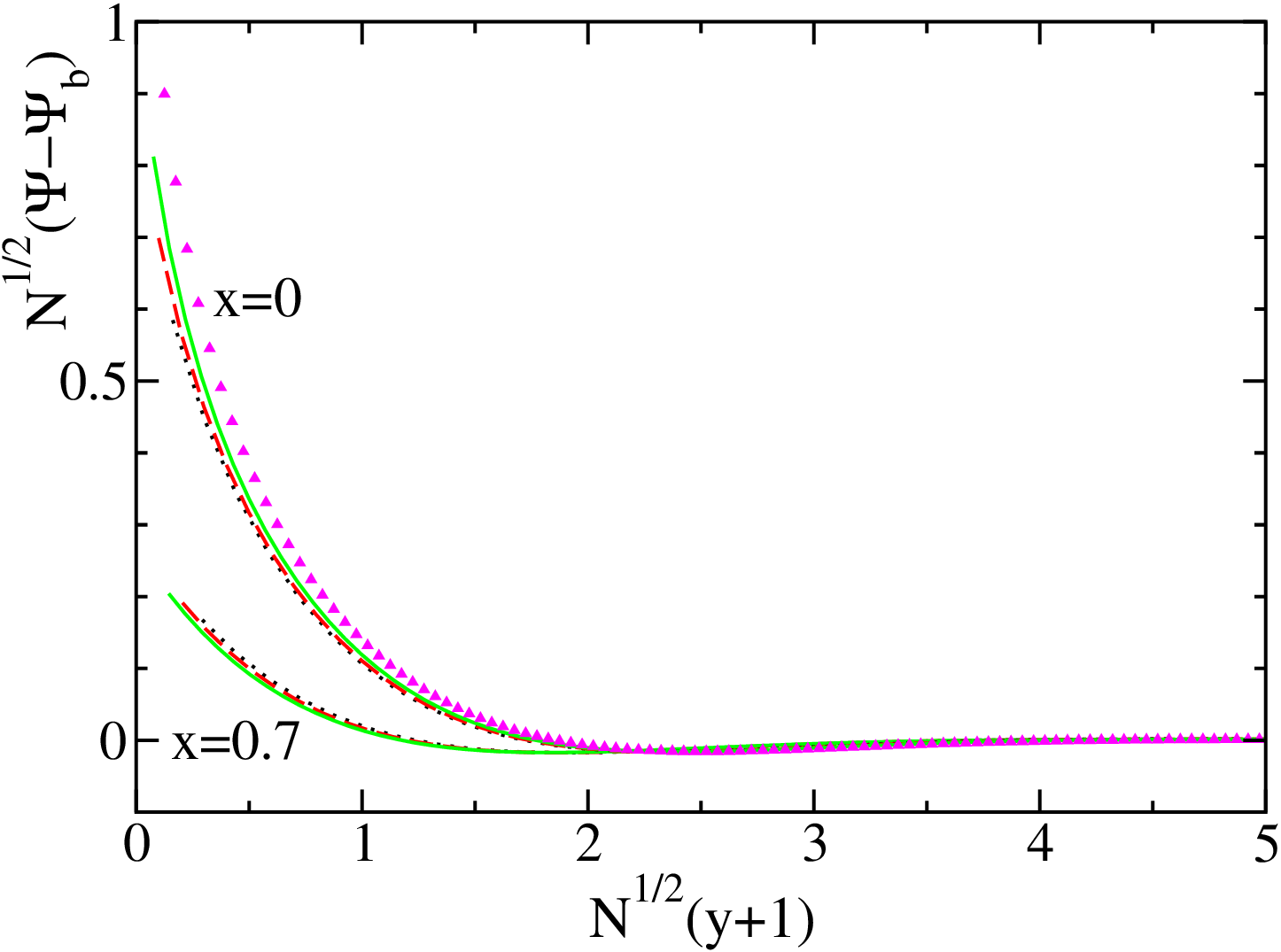}
\caption{Scaling behaviour of  $\psi_{i,j}$ close to the leftmost  edge of the
  chain ($y\approx -1$) for fixed  BC and $\gamma=1$. Dotted, dashed and solid
  lines refer to  $N=200$, $400$, and $800$ and  $\lambda=1$.  The three lower
  (upper)  curves  correspond to  $x=0.7$  ($x=0$). Triangles correspond  to
  $N=1600$, $\lambda=0.25$, and $x=0$.}
\label{blay}
\end{center}
\end{figure}

The variable that is mostly affected by the presence  of such boundary layers
is $\psi$ which even changes its  scaling behaviour with $N$. This is shown in
figure \ref{blay} for the case of  fixed BC. In order to emphasize the scaling
behaviour at the boundary, we subtract from $\psi$ the $1/N$ term (denoted
by $\psi_b$) in the bulk that we know is constant (see \cite{partI}).
For fixed BC and $\lambda=1$, $\psi_b=0$, since the leading term is of order
$1/N^{3/2}$, while for $\lambda=0.25$, $\psi_b = 1/N$ (with a few percent of
uncertainty on the numerical constant.) The data  collapse reveals that
$\psi$ passes from values  of order  $1/\sqrt{N}$ to values of higher order
over a  number of sites of  order $\sqrt{N}$. 

\begin{figure}[!ht]
\begin{center}
\includegraphics[scale=0.5,clip]{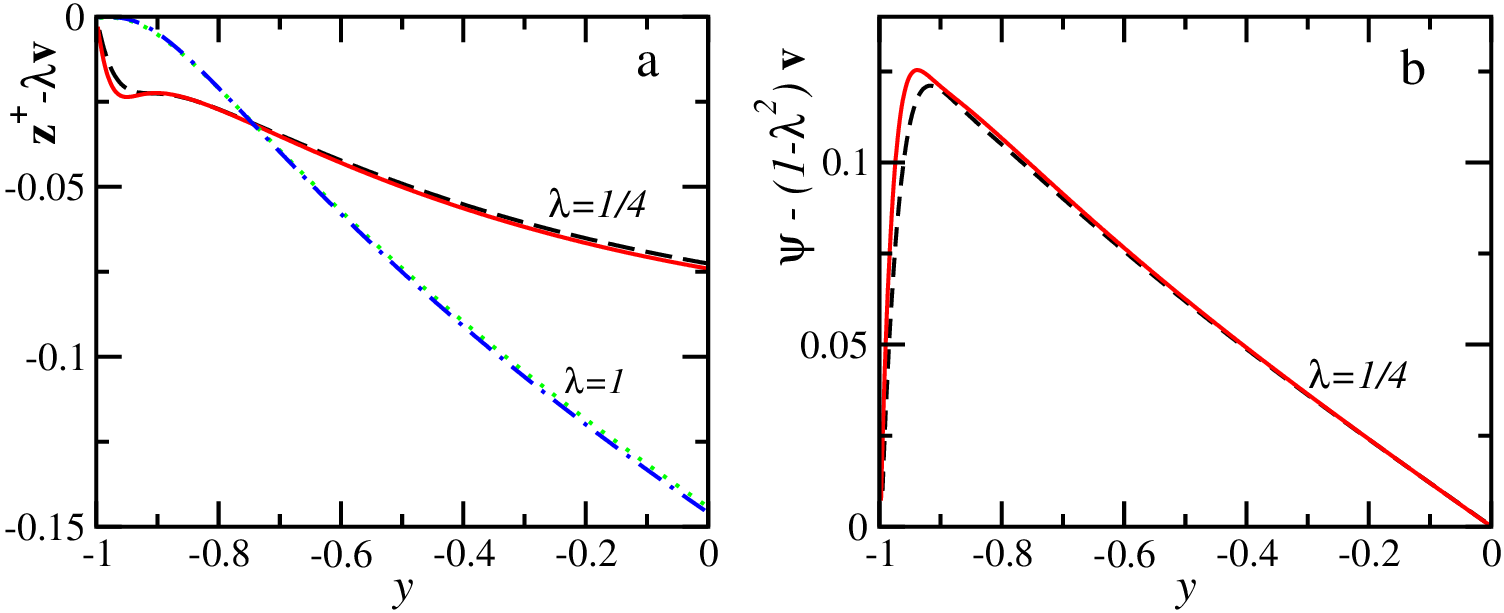}
\caption{The  constraints  \eref{constraints}  among the  correlators  $\psi$,
  $\mZ^+$ and  $\mV$ are plotted as  a function of  $y$ and $\omega=\gamma=1$.
  Dashed and solid curves correspond  to $N=800$ and $1600$, respectively, and
  $\lambda=1/4$.  Dotted  and dotted-dashed  curves correspond to  $N=800$ and
  $1600$, respectively, and $\lambda=1$.}
\label{dif}
\end{center}
\end{figure}

The existence of a  boundary layer  manifests itself in the values that
different correlators assume at the boundaries (in the vicinity of $y=\pm1$).
At the level of the partial differential equations derived in \cite{partI}, the
BC (either free or fixed), lead  to certain  mathematical constraints among the
correlators that must be satisfied for $y=\pm1$. For instance, for free  BC, we
have found analytically that (see Eqs.  (64), (65) and (66) of \cite{partI})
\begin{equ} \label{constraints}
\omega^2\mZ^+(x,-1) - \lambda\mV(x,-1) = 0 \ , \ \textrm{and} \quad (\omega^2 -
\lambda^2)\mV(x,-1) - \omega^2\psi(x,-1) = 0 \ . 
\end{equ}
On the contrary, for  fixed BC we have found that all correlators turn out to
be  zero  at  the  boundaries.   In  figure  \ref{dif}  we  have  plotted  the
combined variables appearing in the l.h.s. of \eref{constraints} for $\omega$
and $\gamma$ unity.  In panel $a$,
we plot  $\mZ^+-\lambda\mV$ as a function  of $y$ for  $\lambda=1$ (dotted and
dotted-dashed curves)  and $\lambda=1/4$ (solid  and dashed curves).   In both
cases,  the combined  variable reaches  zero at  $y\to-1$, thus confirming the
findings in \cite{partI}.  More  importantly, for $\lambda=1/4$, one can infer
that $\mZ^+-\lambda\mV$  will exhibit a  discontinuity at $y=-1$ in  the limit
$N\to\infty$.  This is a direct consequence of the boundary layer which can be
further seen in figure \ref{dif}$b$, where we plot $\psi-(1-\lambda^2)\mV$ for
$\lambda=1/4$.   Furthermore, we  see that  the second  theoretical constraint
\eref{constraints} is also  satisfied.  This means that $\psi$  becomes of the
same order as  $\mV$. From a physical  point of view this implies  that in the
boundary layer,  {\it i.e.}  at a  short distance from the  boundaries (of the
order of $\sqrt{N}$), local equilibrium does not hold.

\section{Relaxation to the stationary state}
\label{sec:relax}

Another interesting issue of nonequilibrium phenomena concerns the convergence
towards  the  stationary  state,  with  a particular  reference  to  the  time
scales. In the context of our stochastic model, this question can be addressed
by investigating time-dependent solutions of equations \eref{eq:Cdot}.  As the
evolution is linear, this can be done by determining the whole spectrum of the
corresponding linear operator.

For computational purposes it is actually convenient to recast the
problem \eref{eq:Cdot} in a more compact way, by a suitable
``unfolding'' of the elements of the matrices $\mY$, $\mV$ and $\mZ$
in a linear array $\mX$. To minimize memory requirements, we take into
account the fact that $\mY$ and $\mV$ are symmetric by construction
and we consider only their independent entries.  On the other hand,
$\mZ$ is antisymmetric only in the stationary state .  Therefore, at
all finite times, all its elements must be considered.  Altogether,
$\mX$ is composed of $M=2N^2+N$ independent elements and the equations
of motion for the correlators can be formally written as
\begin{equation}
\dot \mX\;=\;\L \mX +\mX_0
\label{eqX}
\end{equation} 
where $\L$  is an  $M\!\times\!M$ matrix and  the vector $\mX_0$  contains the
source  terms proportional  to  $\lambda$. The  matrix  $\L$ is  real but  not
symmetric   and    therefore,   has   $M$    complex   conjugate   eigenvalues
$\Lambda=\Lambda_R+i  \Lambda_I$.  Global  stability of  the  stationary state
requires all real parts $\Lambda_R$ to be non positive.

\begin{figure}[!t]
\begin{center}
\includegraphics[clip,scale=0.5]{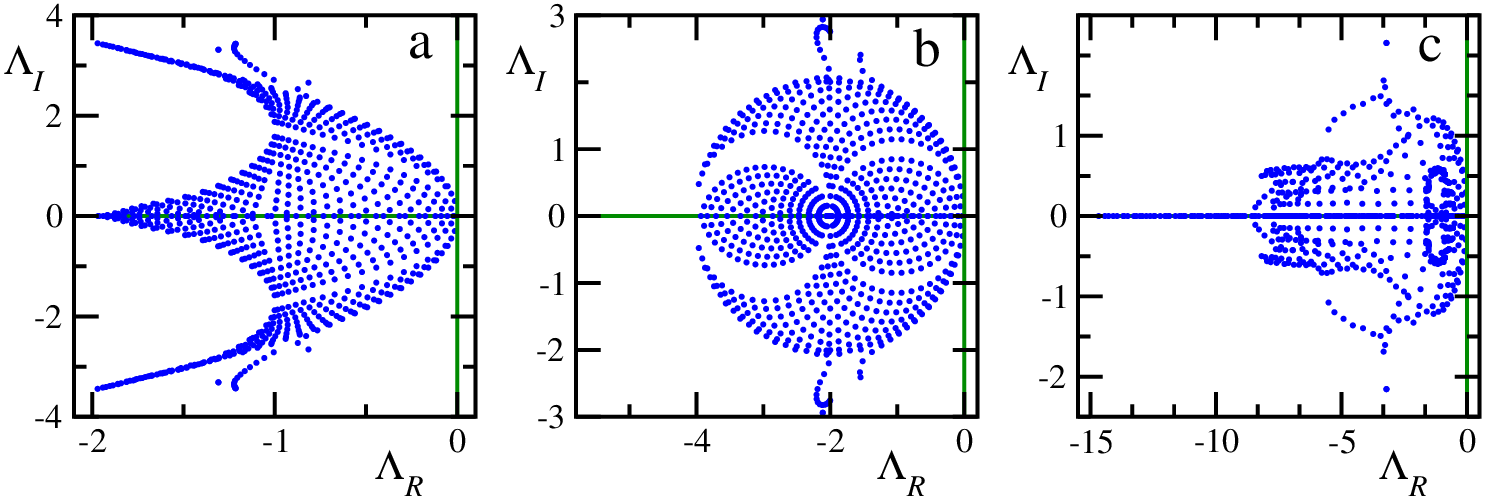}
\caption{Spectra of  the matrix $\L$ for  $N=20$ and $\gamma=0.5,  1,$ and $2$
  (panel $a$,  $b$, and $c$,  respectively). For the  sake of clarity  we have
  used  different scales  along  the  vertical and  horizontal  axes with  the
  exception of  panel $b$, where we wish  to draw the attention  to the nearly
  circular symmetry.}
\label{fig3a}
\end{center}
\end{figure}

The   simplest  approach   consists  in   computing  the   eigenvalue  spectra
$\{\Lambda_i\}$ of  the matrix $\L$  with standard linear  algebra algorithms.
Their location in  the complex plane is illustrated  in figure \ref{fig3a} for
three different values  of $\gamma$.  First we recall  that, to our knowledge,
this is the first nontrivial model where all time scales, from the microscopic
to macroscopic ones  can be obtained at once. As  expected, the whole spectrum
lies  on the negative  $\Lambda_R$ semi-plane,  confirming that  the stationary
state is  stable.  Secondly  we note  that the shape  of the  spectrum changes
qualitatively  upon  varying  the  collision  rate  $\gamma$.   By  increasing
$\gamma$,  the  real  part  of   the  spectrum  is  shifted  towards  negative
values. This is also not surprising as $\gamma$ quantifies the strength of the
internal  stochastic  process  and  thereby of  the  corresponding  relaxation
processes.   More interesting is  the observation  that the  $\Lambda_R$'s are
distributed  over  an entire  range  of scales  from  $\Or(1)$  to very  small
ones. In  the perspective of constructing a  suitable hydrodynamic description
(that is basically the goal of  \cite{partI}), it is only the latter ones that
matter.   Unfortunately,  we  have not  found  a  way  to establish  a  direct
connection between  slow modes (those characterized by  a small $|\Lambda_R|$)
and  hydrodynamic  modes, as  this  would  require  determining not  only  the
eigenvalues, but  also the eigenvectors. This task  is numerically unfeasible,
as the dimension of the space  increases quadratically with $N$, and it is not
even easy to determine the  spectrum, let alone the eigenvectors. In practice,
we have been able to determine the entire spectrum only up to $N \sim 80$.

As  far as  we are  concerned with  the slowest  relaxation processes,  we can
employ an  alternative method  akin to  that used for  the computation  of the
maximum Lyapunov  exponent of a  dynamical system. Indeed,  for asymptotically
long times
\begin{equation}
\mX (t)\;=\;\mX (0) e^{\Lambda_1 t}.
\end{equation}
In order  to estimate $\Lambda_1$, we integrated  numerically the differential
equations (\ref{eqX}) (actually it suffices to consider the homogeneous system
$\dot  \mX\;=\;\L  \mX$ since  $\mX_0$  is  irrelevant)  starting from  random
initial conditions with unit Euclidean  norm, $\|\mX (0)\|=1$. For the sake of
accuracy, we divided the time $T_{tot}$  of the whole run into $n$ consecutive
time intervals, each of length $\tau$,  so that $T_{tot}/\tau = n$. At the end
of each time interval, we  store the corresponding growth rate and renormalize
the vector  $\mX$ to  a unit  norm. Finally, we  determine $\Lambda_1$  as the
average
\begin{equation}
\Lambda_1\;=\; \frac{1}{n\tau}\sum_{l=1}^{n}
\ln \| \mX (l\tau) \|.
\end{equation} 
As a result, we have been able to investigate systems of size up to $N=400$.
\begin{figure}[!t]
\begin{center}
\includegraphics[scale=0.5]{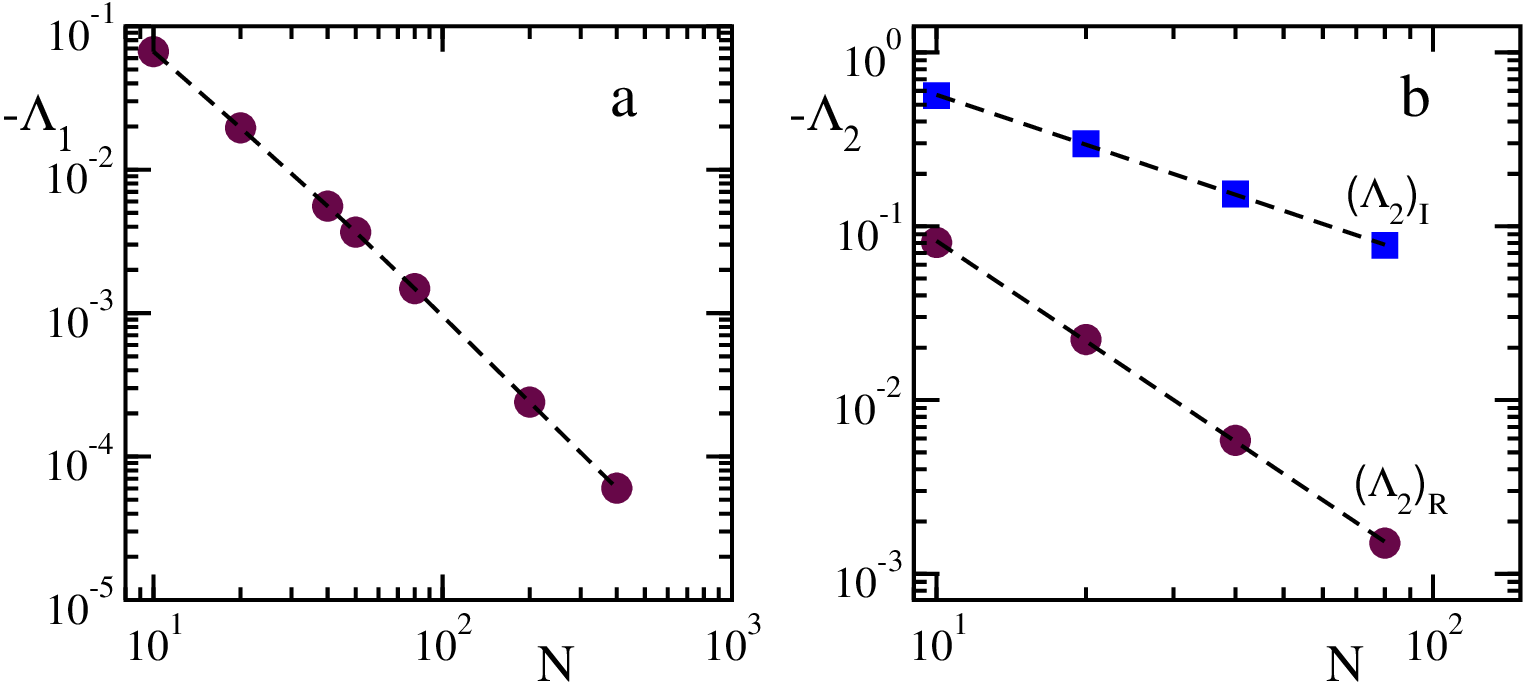}
\caption{Dependence of  the leading eigenvalues  on the chain length  $N$, for
  $\gamma$=1 and fixed  BC. Data is plotted in log-log  scales.  In panel (a),
  the maximum (real) exponent $\Lambda_1$  (circles) is plotted together with a
  power-law  best fit  $N^{-1.91}$  (dashed  line).  In  panel  (b), the  real
  (circles) and imaginary (square)  parts of the second eigenvalue $\Lambda_2$
  are plotted. The  dashed lines correspond to the  power-law fits $N^{-1.91}$
  and $N^{-0.95}$ respectively.}
\label{fig4}
\end{center}
\end{figure}
The numerical results plotted in figure \ref{fig4}a show that for the
considered parameter values, $\Lambda_1$ is real and goes to zero with
some power of $N$.  A best fit suggests that $\Lambda_1 \approx
N^{-2}$. The same approach allows determining the corresponding
(slowest) ``mode'' of the linear operator $\L$.  In figure
\ref{evector} we plot the result obtained for $N=400$.  It looks very
similar to the Fourier modes that we expect on the basis of the
theoretical analysis carried out in \cite{partI} and the order of
magnitude of the corresponding eigenvalue is in agreement with that
analysis.
\begin{figure}[!t]

\begin{center}
\includegraphics[scale=0.65]{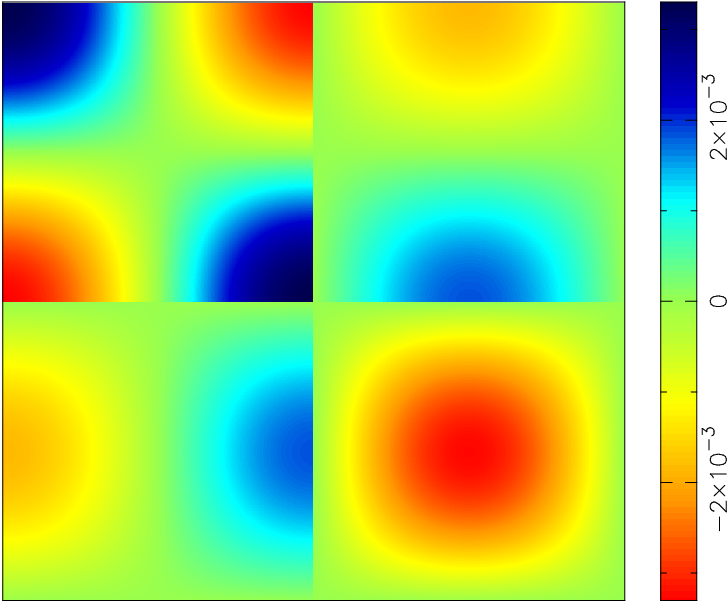}
\caption{Eigenvector of the linear  operator $\L$ corresponding to $\Lambda_1$
  , for  $N=400$ and $\gamma=1$. Starting  from the highest panel  on the left
  and going in clockwise sense, we plot the matrices $\mY_{i,j}$, $\mZ_{i,j}$,
  $\mV_{i,j}$ and $\mZ^{\dag}_{i,j}$ respectively.  }
\label{evector}
\end{center}
\end{figure}

\begin{figure}[!t]
\begin{center}
\includegraphics[clip,scale=0.45,clip]{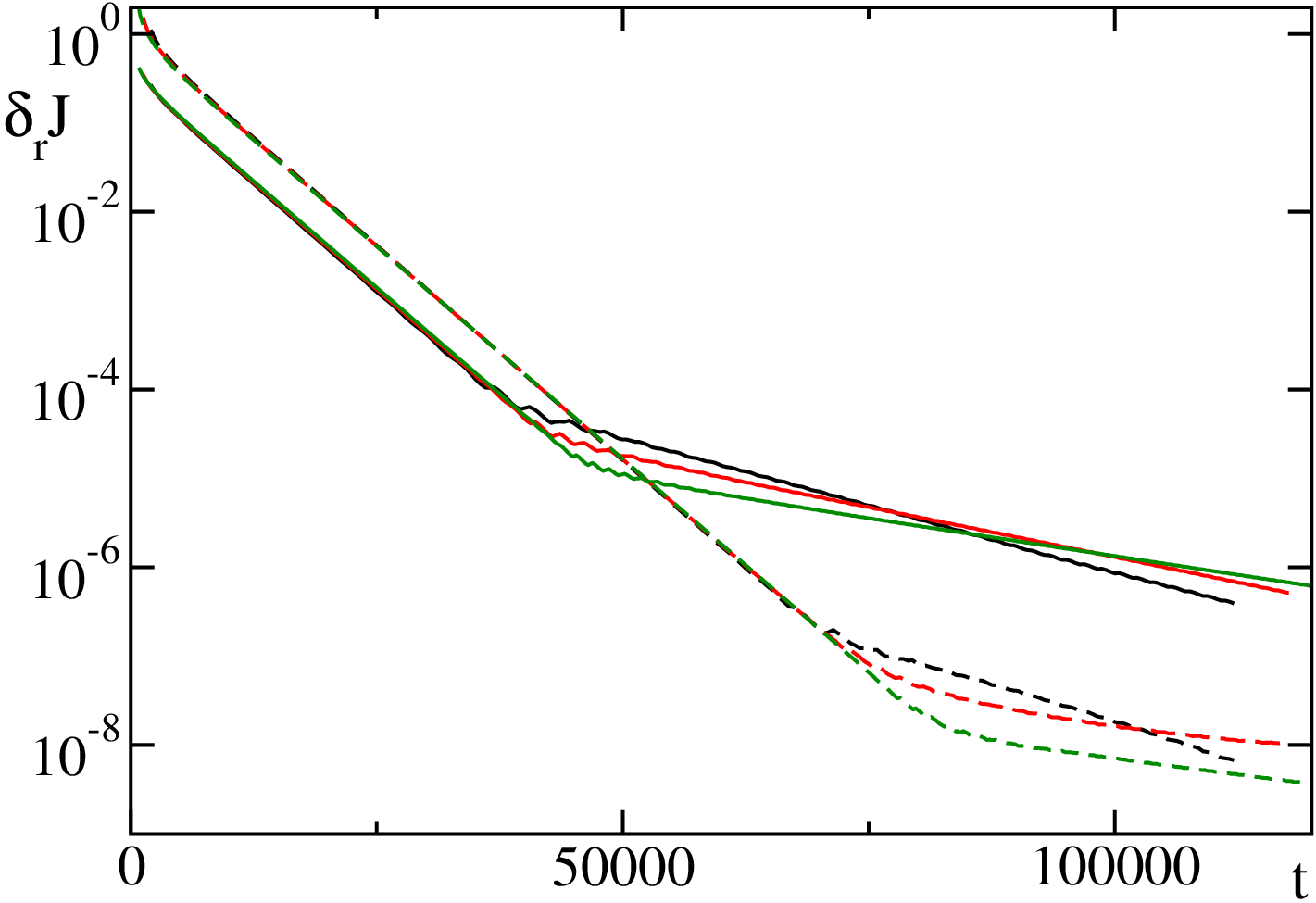}
\caption{Relaxation of the energy flux to its stationary value for fixed BC,
$\gamma=1$ and $\lambda=1$. The relative amplitude of the heat flux (see the
text for its definition) is plotted versus time for three different sizes
($N=200$, 400, 800) and for two definitions of the flux. The dashed curves refer
to the flux along the first bond of the chain; the solid curves refer to the
spatially averaged flux. The time axis has been suitably rescaled to emphasize
the scaling behaviour of the initial exponential regime.}
\label{figf}
\end{center}
\end{figure}

For $N \le  80$ we have been  able to determine the entire  spectrum. It turns
out that  the second and third  eigenvalues are complex  conjugate.  In figure
~\ref{fig4}b we plot  their real and imaginary parts. The  real part scales as
$1/N^{1.91}$, a value that, despite the limited amount of data, suggests again
an asymptotic $1/N^2$ behaviour.  Instead, the imaginary part scales nearly as
$1/N$.  Since this  model is characterized by the presence  of sound waves, we
expect the  imaginary part of $\Lambda_2$  to be connected  to the periodicity
due to the propagation of such  waves.  The period $T$ of the oscillations can
be written as
\begin{equation}
T  \;=\; \frac{2\pi}{(\Lambda_2)_I} \; =  \frac{N}{c} \; ,
\end{equation}
where $c$ is the sound velocity (equal to $\omega$ in our arbitrary units), so
that
\begin{equation}
c\;=\;\frac{ N (\Lambda_2)_I}{2\pi}.
\end{equation}
If we substitute for $(\Lambda_2)_I$ the value found numerically, we obtain $c
\simeq 1$, thus  confirming our expectations.  On the  other hand, the $1/N^2$
dependence of the  real parts poses problems of  consistency with the presence
of an anomalous heat transport, as  the $1/N^2$ dependence is expected to hold
for normal heat  conduction.  In order get a better  understanding of this, we
have investigated  the convergence of  a specific observable, namely  the flux
$J$.  More  precisely, we have studied  the relative deviation of  the flux at
time $t$ from its asymptotic  value (see equation (20) of \cite{LMMP09}).  This
is shown  in figure  \ref{figf} where, starting  from an equilibrium  state at
temperature  $(T_++T_-)/2$  (so  that  $J(0)=0$),  $\delta_r J(t)  \;=\;  1  -
\frac{J(t)}{J(\infty)}$ is plotted as a  function of time.  We have considered
two different  definitions of instantaneous  flux: $i$) the energy  flux along
the  first  bond  (i.e.   directly  in  contact with  the  heat  bath),  which
corresponds to the dashed curves in figure ~\ref{figf}; $ii$) the average flux
(along the  whole chain), which  corresponds to the solid  curves. Altogether,
$\delta_r J(t)$  is a measure  of the deviation  from the stationary  state at
time $t$.   In order  to compare the  curves corresponding to  three different
sizes ($N=200$,  400, and $800$), the  time variable has  been suitably scaled
(by a factor  8.95 for $N=200$ and  2.97 for $N=400$).  The first  part of the
curves  nicely overlap  along a  straight line,  which signals  an exponential
convergence  with a  rate $\eta(N)$  which, taking  into account  the temporal
rescaling factors, is well reproduced by the law
\begin{equation}
\label{eq:fit2}
\eta(N)\;=\;\frac{\eta_0}{N^{3/2}}+\frac{b}{N^{2}}
\end{equation}
A  best fit  of the  numerical  results yields  $\eta_0 \approx  4.4$ in  good
agreement with the second eigenvalue of the operator theoretically derived in
\cite{partI}, that is equal to $4.28$ (see the spectrum plotted in figure 1 
in \cite{partI}, which has been obtained by setting all parameters equal to 1).
The reason why the first eigenvalue does not play any role in our numerical
study is that
the  corresponding eigenmode  is not  excited for  our choice  of  the initial
condition that  is characterized  by exactly the  same average  temperature as
that of the asymptotic stationary state.

The curves reported in figure \ref{figf} show that for any finite $N$
there exists a crossover time beyond which a yet slower convergence
sets in.  By fitting the final slope, one can verify that the time
scale of this last part of the convergence process is on the order of
$N^2$, in agreement with the previous spectral analysis. However, it
is important to notice that this time increases with $N$ and thereby
corresponds to increasingly small scales (look at the vertical axis in
figure \ref{figf}). Altogether, this means that the components of the
initial state that lie along the slowest components become
increasingly small upon increasing the system size $N$, until they
vanish in the thermodynamic limit. One way to understand the
unphysical character of these ``super slow'' modes is as follows. Any
meaningful invariant measure is characterized by a set of correlators,
but the converse is not true.  Only the covariance matrices ${\bf c}$
that are positive semi-definite can correspond to physically
meaningful state.  For instance, we have verified that a sufficiently
large perturbation along the eigenmode depicted in figure
\ref{evector} leads to unphysical matrices.

\section{Discussion and conclusions}
\label{sec:concl}

The study of heat transport in a chain of $N$ particles with
nearest-neighbour coupling and conservative noise allows one to
investigate both analytically \cite{LMMP09} and numerically many
subtle aspects of anomalous transport in one-dimensional systems.
Since an analytical solution is available only for fixed BC, the
free BC case can be investigated only by means of numerical methods.
The comparison between the two cases
shows that the physics of heat transport strongly depends on the
choice of BC. For instance, as already observed in the FPU-$\beta$
model \cite{Comm}, the ratio between the heat fluxes measured with free
and fixed BC does not converge to 1 in the thermodynamic limit ($N \to
\infty$). Moreover, we find that for fixed BC the heat flux and the
temperature profile are independent of the coupling strength with the
thermal bath, while this does not hold for free BC. Nonetheless, the
anomalous scaling of heat conductivity with the system size ($\kappa
\sim N^{1/2} $) is found to be independent of the choice of BC. We
have also investigated the convergence to the stationary state, both
by determining the eigenvalues of the evolution operator of the
covariance matrix, and by following the evolution of the average heat flux
when starting away from the stationary state. The analysis reveals
that over long time scales, the convergence is controlled by a rate
$\eta(N)$ which scales as $N^{-3/2}$. This means that if one wishes to
extract reliable numerical data by performing direct numerical
simulations, e.g., in a lattice of size $N=50000$, it is necessary to
evolve the system well above $10^7$ time units. It should be kept in
mind that similar limitations hold for deterministic nonlinear
systems, even though they are, in general, characterized by slightly
different exponents \cite{Comm}.  Finally, our analysis of the time-dependent
solution has revealed a crossover from a typical fractional diffusion regime
to a superslow relaxation. The crossover time is found to increase with
the system size, suggesting that the latter regime becomes irrelevant in the
thermodynamic limit.

\ack We acknowledge  Gianpiero Puccioni for his support  in the implementation
of the numerical codes. This work is partially supported by the the Italian
project {\it Dinamiche cooperative in strutture quasi uni-dimensionali} No. 827
within the CNR program Ricerca spontanea a tema libero. C.M.-M. acknowledges 
partial funding from the European Research Council and the Academy of Finland.

\section*{References}
\bibliographystyle{unsrt}
%\bibliography{dupalnum}

\end{document}